\documentclass[aps,twocolumn,groupedaddress,showpacs]{revtex4}
\usepackage{graphicx}

\begin{document}
%%%%%%%%%%%%%%%%%%%%%%%%%%%%%%%%%%%%%%%%%%%%%%%%%%%%%%%%%%%%%%%%%%%%%%%%%%%%%
% You should use BibTeX and revtex.bst for references
%\bibliographystyle{apsrev}
%%%%%%%%%%%%%%%%%%%%%%%%%%%%%%%%%%%%%%%%%%%%%%%%%%%%%%%%%%%%%%%%%%%%%%%%%%%%%
% marks overfull lines with blackboxes
%\draft - no longer supported, use the 'draft' option instead
%\preprint{}
\title{Supersolid of Hardcore Bosons on the Face Centered Cubic Lattice}
% Optional argument for running titles on pages
%\title[]{}
\author{Takahumi Suzuki}
%\thanks{}
%\altaffiliation[Present Address: ]{ISSP, University of Tokyo, Kashiwanoha 5-1-5, Kashiwa, Chiba 277-8581, Japan}
%\affiliation{Department of Applied Physics, Osaka University, Suita, Osaka 565-0871, Japan}
\author{Naoki Kawashima}
\affiliation{Institute for Solid State Physics, University of Tokyo, Kashiwa, Chiba 277-8581}
\date{\today}
%%%%%%%%%%%%%%%%%%%%%%%%%%%%%%%%%%%%%%%%%%%%%%%%%%%%%%%%%%%%%%%%%%%%%%%%%%%%%
%                         ABSTRACT                                          %
%%%%%%%%%%%%%%%%%%%%%%%%%%%%%%%%%%%%%%%%%%%%%%%%%%%%%%%%%%%%%%%%%%%%%%%%%%%%%
\begin{abstract}
We investigate a supersolid state in hardcore boson models on the face-centered-cubic (FCC) lattice. The supersolid state is characterized by a coexistence of crystalline order and superfluidity. Using a quantum Monte Carlo method based on the directed-loop algorithm, we calculate static structure factors and superfluid density at finite temperature, from which we obtain the phase diagram. The supersolid phase exists at intermediate fillings between a three-quarter-filled solid phase and a half-filled solid phase. We also discuss the mechanism of the supersolid state on the FCC lattice.
\end{abstract}
%%%%%%%%%%%%%%%%%%%%%%%%%%%%%%%%%%%%%%%%%%%%%%%%%%%%%%%%%%%%%%%%%%%%%%%%%%%%%
%\maketitle must follow title, authors, abstract and \pacs
\maketitle
%\section{}
%\label{}
%\subsection{}
%\subsubsection{}
%%%%%%%%%%%%%%%%%%%%%%%%%%%%%%%%%%%%%%%%%%%%%%%%%%%%%%%%%%%%%%%%%%%%%%%%%%%%%
%                        MAIN TEXT                                          %
%%%%%%%%%%%%%%%%%%%%%%%%%%%%%%%%%%%%%%%%%%%%%%%%%%%%%%%%%%%%%%%%%%%%%%%%%%%%%

Whether a supersolid state, where both of solidity and superfluidity coexists, realizes in matters or not. The possibility of the supersolid in ${\rm ^4He}$ have been discussed theoretically\cite{Andreev,Chester,Leggett,Matsuda,Liu}. Recently, characteristic behaviors of superfluidity have been reported in torsional oscillator experiments on the solid ${\rm ^4He}$ by E. Kim and M. H. W. Chan(KC)\cite{Kim}, while a number of experiments failed to detect the superfluidity in the solid ${\rm ^4He}$. In their measurements, a sudden drop in the resonant period was observed around $T \sim$ 0.2K. The drop implies emergence of non-classical rotational inertia in the solid ${\rm ^4He}$. They concluded that this was a signature of a transition into the supersolid phase. However, some experimental and theoretical studies suggested different interpretations of KC's observation. For instance, the signal of the superfluidity in the solid ${\rm ^4He}$ became weaker as annealing cycles were repeated\cite{Ritter} and the superflow was blocked by the solid ${\rm ^4He}$ with no grain boundaries\cite{sasaki}. Theoretically, the possibility of the superflow induced by vacancies in the commensurate solid ${\rm ^4He}$, in which the total number of atoms equals a multiple of the number of lattice sites, was ruled out\cite{Clark,Boninsegni}, and the superglass behavior appeared in a quenched system\cite{Boninsegni2}. Although several mechanisms of the superflow in the solid ${\rm ^4He}$ have been proposed, the satisfactory interpretation of KC's results is still controversial.

Bosonic lattice model was introduced as a reasonable model of liquid ${\rm ^4He}$\cite{MM}. Recently, the possibility of the supersolid on the lattice model in triangular lattice\cite{Wessel,Heidarian,Melko,Boninsegni3} and kagome lattice\cite{kagome} cases was studied by quantum Monte Carlo simulations. From these studies it became clear that the frustrated interactions on the triangular lattice stabilize the super current induced by vacancies in the crystalline ordering with the wave vector ${\bf Q}$=$(4\pi/3,0)$ or $(2\pi/3,0)$. In contrast, the supersolid is not stabilized on the kagome lattice where the frustrated interactions exist. In the three dimensional lattice cases, the phase diagrams of the system on the body-centered-cubic (BCC) lattice was obtained by a mean-field approximation and concluded that the supersolid state appears if the next-nearest-neighbor interactions are present\cite{Matsuda,Liu}. However, the reason for the stabilization of the supersolid state on the BCC lattice and the microscopic picture was not cleared from the mean-field results. (Note that the BCC lattice is bipartite and has no frustration if one does not take into account the next-nearest-neighbor interactions.)

Theoretical study beyond the mean-field theory for three dimensional systems is still missing. In this letter, the supersolid state in a three-dimensional bosonic-lattice model is studied by a quantum Monte Carlo method based on the directed loop algorithm\cite{Sandvik,Harada}. We wish to address a generic question what ingredient is necessary to realize the supersolidity. From the study of the two-dimensional case mentioned above, it is presumable that the geometrical frustration plays an essential role in the supersolidity in the bosonic lattice model. We therefore focus on a hardcore-bosonic model on the face-centered-cubic (FCC) lattice, which does not have a direct connection to the lattice structure of the real solid helium, because the FCC lattice is one of the simplest lattices with geometric frustration.

More specifically, we consider bosonic lattice model with the positive hopping amplitude $t>0$ and the nearest-neighbor repulsion $V>0$ on the FCC lattice. The model Hamiltonian is defined by
\begin{eqnarray}
{\mathcal H}=-t\sum_{\langle ij \rangle} \left( {b_{i}}^{\dagger}b_{j} + h.c. \right) + V\sum_{\langle ij \rangle}\hat{n}_{i}\hat{n}_{j} -\mu\sum_{i}\hat{n}_{i},
\label{Ham1}
\end{eqnarray}
where $\mu$ is the chemical potential, ${b_{i}}^{\dagger}(b_{i})$ is the bosonic creation (annihilation) operator, and $\hat{n}_{i}={b_{i}}^{\dagger}b_{i}$. The summation $\langle ij \rangle$ is over the nearest neighbor pairs and the system size is defined by $N=L^3$. The periodic boundary condition is applied. Under the hardcore condition, the original bosonic-lattice model is identically mapped onto the $S$=$1/2$ XXZ model,
\begin{eqnarray}
{\mathcal H}&=&-J_{\perp}\sum_{\langle ij \rangle}({S_{i}}^{x}{S_{j}}^{x}+{S_{i}}^{y}{S_{j}}^{y})-J_{z}\sum_{\langle ij \rangle}{S_{i}}^{z}{S_{j}}^{z}\nonumber\\
& &-H\sum_{i}{S_{i}}^{z},
\label{Ham2}
\end{eqnarray}
where $J_{\perp}=2t$, $J_z=-V$ and $H=\mu-6V$. Note that $J_\perp$ and $J_z <(>)0$ mean the antiferromagnetic (ferromagnetic) interactions. In the spin language, the supersolidity is characterized by the following two properties: sublattice-dependent expectation values of the longitudinal spin components (broken translational symmetry, or crystallization), and non-vanishing transverse spin components (off-diagonal long range order, or superfluidity).

\begin{figure}[htb]
\begin{center}
\includegraphics[trim=5mm 0mm 100mm 0mm,angle=270, scale=0.33]{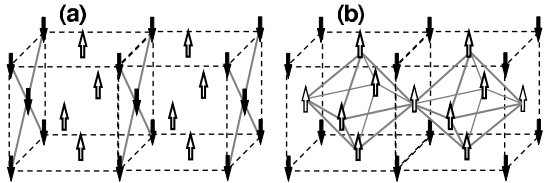}
\vspace{-5 mm}
\caption{The perfectly ordered spin configurations. (a) ${\rm AB}$ state and (b) ${\rm A_3B}$ state. Note that the dashed lines connecting next-nearest neighbors are mere guide lines to the eye and there is no direct coupling corresponding to these lines. (The direct couplings exist only for nearest neighbor pairs.) The gray lines denote the superflow paths in the supersolid state (see text).}
\label{spin-arr}
\end{center}
\end{figure}

In the limit $J_{\perp}/|J_{z}|\rightarrow0$ (Ising model), the ordered states were investigated and the $H$-$T$ phase diagram was obtained\cite{Binder,Hagai,Kammerer}. At the magnetization $m$=$\langle\sum_i {S_{i}}^z/N \rangle$=$0$ and $1/4$, there appear two solid phases conventionally referred to as ${\rm AB}$ and ${\rm A_3B}$. Representative spin configurations of the two phases are shown in Fig. \ref{spin-arr}. The phase transition from the ${\rm AB}$ phase to the ${\rm A_3B}$ phase occurs at $H_{\rm Ising}=2|J_z|$ at absolute zero temperature.

\vspace{-5 mm}

In order to investigate the crystalline order and the off-diagonal long-range order for $J_{\perp}> 0$, we calculate the static structure factor $S({\bf Q})$ (SSF) and the superfluid density $\rho_s$\cite{Pollack}, defined by
\begin{eqnarray}
S({\bf Q})&=&\left\langle \left| \sum_{i} \exp [i{\bf Q}\cdot{\bf r}_{i}] S_{i}^{z} \right|^{2}\right\rangle
\label{sq}
\end{eqnarray}
and
\begin{eqnarray}
\rho_s&=&\frac{k_BT\left\langle {\bf W}^2 \right\rangle}{3J_{\perp}L},
\end{eqnarray}
where ${\bf W}=(W_x,W_y,W_z)$ denotes the winding number of the world-lines. In what follows, we express the wave vector ${\bf Q}$ by the conventional choice of the unit reciprocal vectors. 

When the system is in the ${\rm AB}$ or ${\rm A_3B}$ ordered state, the SSF is proportional to $N$ and strong system-size dependence is expected at ${\bf Q_{sol}}$=$(\pi,\pi,0)$, $(\pi,0,\pi)$, and $(0,\pi,\pi)$. For other ${\bf Q}$s, the SSF should be system-size independent. It was reported for the classical case ($J_{\perp}$=$0$)\cite{Kammerer,Ackermann} that a perfect solid state, either ${\rm AB}$ or ${\rm A_3B}$, can hardly be observed in a system of computationally accessible size due to antiphase domain boundaries (APB). While the states with domain boundaries have negligible weight in the thermodynamic limit, they have non-negligible contributions for small systems because the domain-wall free-energy is small due to the frustrated nature of the interactions. The APBs reduce $S({\bf Q_{sol}})$ since contributions from different phases have opposite signs. However, the magnitude of the reduction depends on the locations of the APBs and the cancellation does not in general make $S({\bf Q_{sol}})$ completely vanishing. Therefore, the average $S({\bf Q_{sol}})$ is still proportional to the system size even if the APBs are present. As for the effect of the APBs on the superfluid density, we have confirmed that it is relatively minor compared to that on $S({\bf Q_{sol}})$. To see this, we evaluated $\rho_s$ in two ways (see Fig. 4 (b)). One is a long equilibrium simulation starting from random initial configurations, in which APBs are observed. The other is relatively short Monte Carlo simulations starting from the perfect ${\rm AB}$ or ${\rm A_3B}$ configuration. In the latter, the length of the simulation is chosen such that APBs do not appear. In both cases, the superfluid density yielded the same value within the statistical error. 

\begin{figure}[htb]
\begin{center}
\includegraphics[trim=0mm 0mm 0mm 0mm ,scale=0.38]{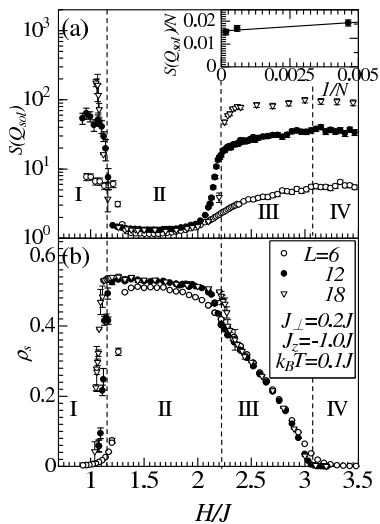}
\vspace{-3 mm}
\caption{The field dependence of $S({\bf Q_{sol}})$ and $\rho_s$ at $(J_{\perp},J_z)=(0.2,-1.0)J$ and $k_BT=0.1J$. The open circles, the solid circles, and the inverted triangles denote the results of $L=6$, $12$, and $18$, respectively. Note that the results of $S({\bf Q_{sol}})$ are averaged values of $S(0,\pi,\pi)$, $S(\pi,0,\pi)$ and $S(\pi,\pi,0)$. The inset shows the system size dependence of $S({\bf Q_{sol}})/N$ at $H/J_z=3.0$.}
\label{sq-res}
\end{center}
\end{figure}

In Fig. \ref{sq-res}, we show the results of the field dependence of $S({\bf Q_{sol}})$ and $\rho_s$ at $(J_{\perp},J_z)=(0.2,-1.0)J$ and $k_BT=0.1J$, where $J\equiv|J_z|$ is our unit of energy. The $H$-axis can be divided into four regions according to the behaviors of $S({\bf Q_{sol}})$ and $\rho_s$; (I) the low-field region $0<H<H_{\rm solid1}\sim1.15J$, (II) the lower-intermediate region $H_{\rm solid1}<H<H^*\sim2.2J$, (III) the upper-intermediate region $H^*<H<H_{\rm solid2}\sim3.1J$, and (IV) the high-field region $H_{\rm solid2}<H$, where $H_{\rm solid1}$, $H^*$, and $H_{\rm solid2}$ are temperature-dependent transition fields. 

\begin{figure}[htb]
\begin{center}
\includegraphics[trim=0mm 0mm 0mm 0mm ,scale=0.40]{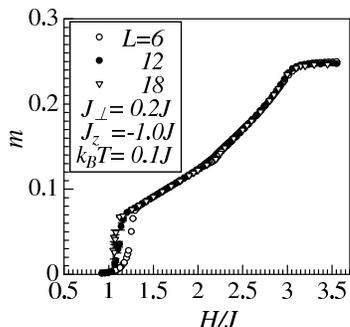}
\vspace{-5 mm}
\caption{The field dependence of the magnetization at $(J_{\perp},J_z)$=$(0.2, -1.0)J$ and $k_BT$=$0.1J$. The value of $m$=$1/4$ corresponds to a half of the saturation magnetization.}
\label{sz-res}
\end{center}
\end{figure}

In the regions I, III and IV, the crystalline order exists. This is evident from the fact that $S({\bf Q_{sol}})$ increases in proportion to the system size. In the region II, on the other hand, $S({\bf Q_{sol}})$ does not show a system-size dependence indicating no crystalline ordering in this region. The superfluid density $\rho_s$ is almost zero in the regions I and IV whereas in the intermediate regions II and III, it stays finite. Judging from these results, we conclude that the ground state is the solid state in the regions I and IV. As shown in Fig. \ref{sz-res}, the magnetization plateaus at $m$=$0$ and $1/4$ appear in the corresponding fields. Therefore, these solid phases are the ${\rm AB}$ and ${\rm A_3B}$ ordered phases, respectively. The region II is the superfluid phase. Finally in the region III, since the crystalline order and the superfluidity coexist, there appears the supersolid phase. Hence, we conclude that the supersolid state is stable in three dimensions.


\begin{figure}[htb]
\begin{center} 
\includegraphics[trim=0mm 0mm 0mm 0mm ,scale=0.38]{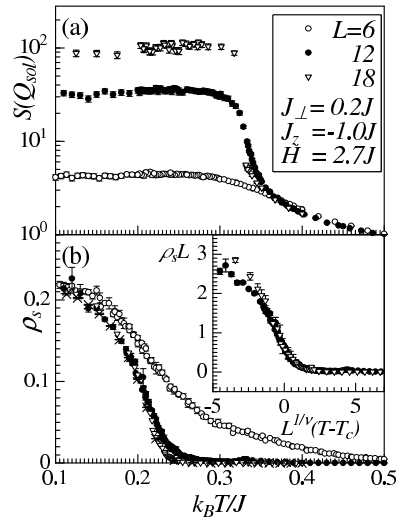}
\vspace{-3 mm}
\caption{The temperature dependence of $S({\bf Q_{sol}})$ and $\rho_s$ for $(J_{\perp},J_z)$=$(0.2, -1.0)J$ and $H$=$2.7J$. In (b), the cross symbols denote the results starting from the perfect ${\rm A_3B}$ configuration in $L$=$18$ and the others are those starting from the random initial configurations. The inset in (b) is the finite-size scaling of the superfluid density.}
\label{scaling}
\end{center}
\end{figure}

\begin{figure}[htb]
\begin{center}
\includegraphics[trim=20mm 140mm 50mm 15mm,scale=0.45]{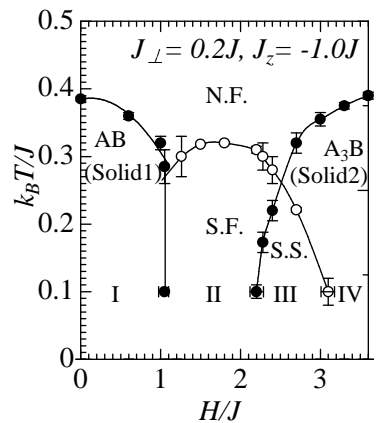}
%\vspace{-3 mm}
\caption{The $H$-$T$ phase diagram for $(J_{\perp},J_z)$=$(0.2,-1.0)J$. The open circles and the solid circles indicate the first-order and the second-order transitions, respectively. The solid lines are mere guide lines to the eye. The labels, "N.F.", "S.F.", and "S.S." stand for the normal fluid phase, the superfluid phase, and the supersolid phase, respectively.}
\label{phased}
\end{center}
\end{figure}

Next we study the temperature dependence. The results at $H=2.7J$ (in the supersolid region III) are shown in Fig. \ref{scaling}. As we decrease the temperature with fixed magnetic field, $S({\bf Q_{sol}})$ almost discontinuously increases at $k_BT$=$k_BT_{{\rm solid}}\sim0.32J$ and the system-size dependence appears for $T<T_{{\rm solid}}$. However, the superfluid density remains very small; $\rho_s<5\times 10^{-3}$ near $T_{{\rm solid}}$. This is the transition from the normal fluid to the solid phase. Since $S({\bf Q_{sol}})$ takes the same values of those in the ${\rm A_3B}$ solid region IV and scarcely shows the temperature dependence in $T<T_{\rm solid}$, we identify this solid phase as the ${\rm A_3B}$ ordered phase. At even lower temperature, the system undergoes another transition from the solid phase to the supersolid phase. This is marked by the increase in the superfluid density $\rho_s$ that starts at $k_BT_{{\rm super}}\sim0.22J$. To estimate $T_{{\rm super}}$, we analyze the finite-size-scaling behavior of the superfluid density $\rho_s$ by the scaling form $\rho_s L$=$f(L^{1/\nu}(T-T_{\rm super}))$ using the exponents of the three dimensional $XY$ model $\nu$=$0.6723$\cite{Hasenbusch}. As shown in the inset of Fig. \ref{scaling}(b), the data collapse is obtained with the critical temperature, $k_BT_{{\rm super}}$=$0.221(2)J$. Thus, we conclude that the phase transition from the solid phase to the supersolid phase is of the second order and its universality class is that of the three dimensional $XY$ model as expected. In this way, we estimate the critical temperatures $T_{{\rm solid}}$ and $T_{{\rm super}}$ for various other values of $H$ to obtain the phase boundary. Here, $T_{{\rm solid}}$ and $T_{{\rm super}}$ denote the transition temperatures where the crystalline order and the superfluid order, respectively, emerge. The results are shown in Fig. \ref{phased}.


From the phase diagram in Fig.\ref{phased}, we find that the supersolid phase exists between the ${\rm AB}$ and ${\rm A_3B}$ solid phases. This region locates slightly above $H_{\rm Ising}$=$2.0J$ at which the phase transition occurs in the classical case ($J_{\perp}$=$0$) from the ${\rm AB}$ to ${\rm A_3B}$ phase. To discuss the mechanism of the supersolid state on the FCC lattice, we consider the spin configuration above $H_{\rm Ising}$. In the classical case, at the critical field, the spins at the centers of faces of the cubic lattice (the up spins in Fig. \ref{spin-arr}(b)) become {\it dangling} spins; they can be reversed without changing the energy. Let us regard down (up) spins at these locations as hard-core particles (holes). Then, $H-H_{\rm Ising}$ can be interpreted as the excitation gap for creating a particle and the ground state is the empty state at $H$ larger than the critical value. However, once the hopping ($J_{\perp}$) is turned on, the excited particles may move along the gray lines in Fig.\ref{spin-arr}(b) and can in general condense. If the magnetic field is far larger than $H_{\rm Ising}$, the classical gap is larger than the scale of the hopping constant $J_{\perp}$ (i.e., the band width of particle excitation), the gap remains open even if the quantum hopping is present. As we decrease the magnetic field, however, at some point the classical gap becomes smaller than the scale of the hopping constant. Accordingly the actual gap closes and the ground state starts exhibiting superfluidity. At this point, in contrast to the spins on the faces, the spins at the corners can hardly be affected by the hopping term, because the energy cost of reversing one of these spins is $\Delta E$$\sim\frac{3}{4}|J_z|$ and is still too large. Therefore, they stay in a solid crystalline order. As we further decrease the magnetic field, the density of condensed particles at the dangling spin locations gradually increases. This generates positive molecular fields at the corners, destabilizing the crystalline order. This destabilizing effect finally melts the crystal at $H$=$H^{\ast}$. This latter transition point must be larger than $H_{\rm Ising}$ because the transition must take place before the classical excitation gap closes and therefore the density of the excited particles diverges. This is the microscopic scenario of the two transitions, the solid to the supersolid transition and the supersolid to the super fluid transition. Indeed, we successfully confirmed the existence of the ${\rm A_3B}$-type supersolid states in the corresponding parameter region in the present simulation.

This scenario predicts a supersolid phase of another type in the region $H<H_{\rm Ising}$, which we could call ${\rm AB}$-type. The mechanism of the ${\rm AB}$-type supersolid is again understood by the {\it dangling} spins. This time, the dangling spins appear at the sites occupied by down spins in Fig.\ref{spin-arr}(a), and we should regard the up spins on these sites as excited particles, which condense in the AB-type supersolid phase that locates below $H_{\rm Ising}$. The excited particles hop along the gray lines in Fig.\ref{spin-arr}(a), while rigid spins (those pointed up in Fig.\ref{spin-arr}(a)) stay in the crystalline order. The ${\rm AB}$-type supersolid has the characteristic two-dimensional paths of the superfluid due to the alternatively stacks of the superfluid and solid layers, while the ${\rm A_3B}$-type supersolid has the three dimensional superfluid connections. Perturbatively, the effective interactions between these superfluid layers may arise in the second order of $J_{\perp}$. In Fig. 6, we show some results of the superfluid density and the structure factor at $(J_{\perp},J_z)$=$(0.15,-1.0)J$ and $H$=$1.65J<H_{\rm Ising}$. In this case, while an anomaly, which is cleared by the calculations in the larger system size, appears in $S({\bf Q_sol})$ at $k_BT\sim 1.5$ due to the APB's, the ${\rm AB}$-type supersolid realizes in the region $k_{B}T < k_{B}T_{\rm super} \sim 0.104(2)$.

\begin{figure}[htb]
\begin{center}
\vspace{3 mm}
\includegraphics[trim=0mm 0mm 0mm 0mm ,scale=0.35, angle=270]{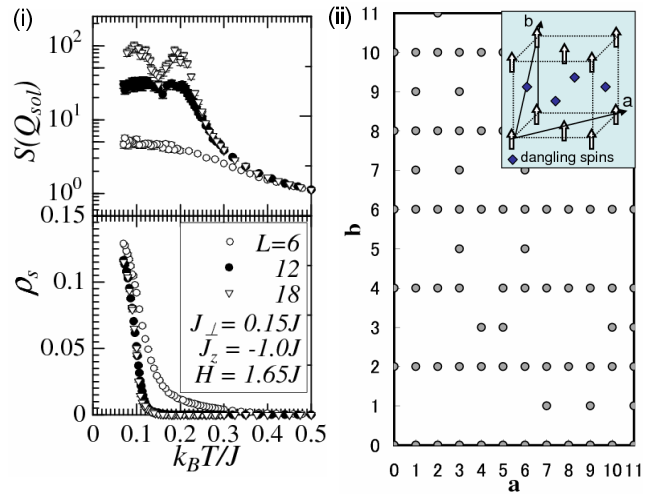}
\vspace{-3 mm}
\caption{(i)The temperature dependence of $S({\bf Q_{sol}})$ and $\rho_s$ in $(J_{\perp},J_z)$=$(0.15, -1.0)J$ and $H$=$1.65J$. (ii)The cross-section of a spin configuration in the AB-type supersolid at $k_BT=0.1$. The gray circles denote up spins and the unoccupied sites correspond to down spins.}
\label{ABss}
\end{center}
\end{figure}

To summarize, we have calculated the SSF and $\rho_s$ for $S$=$1/2$ XXZ model on the FCC lattice and obtained a phase diagram at fixed $J_{\perp}/J_z$. We have also discussed the microscopic mechanism of the supersolidity in the present model and pointed out that {\it the connections of dangling spins} resulting from the geometrical frustration play a key role in the formation of the supersolid state.

We would like thank Y. S. Wu, M. Kohmoto, K. Harada, and C. Batista for useful comments and fruitful discussions. This work is supported by Next Generation Supercomputing Project, Nanoscience Program, MEXT, Japan. Numerical computations were carried out at the facilities of the Supercomputer Center, Institute for Solid State Physics, University of Tokyo.

%%%%%%%%%%%%%%%%%%%%%%%%%%%%%%%%%%%%%%%%%%%%%%%%%%%%%%%%%%%%%%%%%%%%%%%%%%%%%
%                            REFERENCES                                     %
%%%%%%%%%%%%%%%%%%%%%%%%%%%%%%%%%%%%%%%%%%%%%%%%%%%%%%%%%%%%%%%%%%%%%%%%%%%%%
% Create the reference section using BibTeX
%\bibliography{nmr}
%%%%%%%%%%%%%%%%%%%%%%%%%%%%%%%%%%%%%%%%%%%%%%%%%%%%%%%%%%%%%%%%%%%%%%%%%%%%%

%
\end{document}